\documentclass[twocolumn]{aastex631}
\usepackage{hyperref}
\usepackage{CJKutf8}

\newcommand{\lycl}{\rm{LyC\ leakers}}
\newcommand{\ionone}{\emph{Ion1}}
\newcommand{\iontwo}{\emph{Ion2}}
\newcommand{\cdfs}{\emph{CDFS-6664}}

\begin{document}

\title{Lyman Continuum Leakers at $z>3$ in the GOODS-S Field: Starburst or Not?}

\begin{CJK*}{UTF8}{gbsn}
\author[0000-0002-2528-0761]{Shuairu Zhu（朱帅儒）}
\affiliation{Shanghai Astronomical Observatory, Chinese Academy of Sciences, 80 Nandan Road, Shanghai 200030, People’s Republic of China}
\affiliation{School of Astronomy and Space Sciences, University of Chinese Academy of Sciences, No. 19A Yuquan Road, Beijing 100049, People’s Republic of China}

\author[0000-0001-6763-5869]{Fang-Ting Yuan*}
\affiliation{Shanghai Astronomical Observatory, Chinese Academy of Sciences, 80 Nandan Road, Shanghai 200030, People’s Republic of China}
\correspondingauthor{Fang-Ting Yuan}
\email{*email: yuanft@shao.ac.cn}

\author[0000-0002-0003-8557]{Chunyan Jiang}
\affiliation{Shanghai Astronomical Observatory, Chinese Academy of Sciences, 80 Nandan Road, Shanghai 200030, People’s Republic of China}

\author[0000-0002-9634-2923]{Zhen-Ya Zheng}
\affiliation{Shanghai Astronomical Observatory, Chinese Academy of Sciences, 80 Nandan Road, Shanghai 200030, People’s Republic of China}
\affiliation{School of Astronomy and Space Sciences, University of Chinese Academy of Sciences, No. 19A Yuquan Road, Beijing 100049, People’s Republic of China}

\author[0000-0003-3987-0858]{Ruqiu Lin}
\affiliation{Shanghai Astronomical Observatory, Chinese Academy of Sciences, 80 Nandan Road, Shanghai 200030, People’s Republic of China}
\affiliation{School of Astronomy and Space Sciences, University of Chinese Academy of Sciences, No. 19A Yuquan Road, Beijing 100049, People’s Republic of China}

\defcitealias{Vanzella2012}{V12}
\defcitealias{Gupta2024}{G24}
\defcitealias{Ji2020}{J20}
\defcitealias{Debarros2016}{DB16}
\defcitealias{Yuan2021}{Y21}
\defcitealias{Saxena2022}{S22}
\defcitealias{Rivera-thorsen2022}{RT22}
\defcitealias{Bruzual2003}{BC03}
\defcitealias{Kerutt2024}{K24}

\begin{abstract}
We investigate the star-forming properties of 23 Lyman Continuum (LyC) leakers at $z>3$ in the Great Observatories' Deep Survey-South (GOODS-S) field based on a systematic review of LyC observations from the literature. 
Using data from the Hubble Space Telescope (HST) and the James Webb Space Telescope (JWST), we construct the spectral energy distributions (SEDs) for these LyC leakers, covering the spectrum from rest-frame ultraviolet to near-infrared. Through the application of a unified modeling approach, we measure the ultraviolet slope ($\beta$), star formation rate, and stellar mass for these LyC leakers in a consistent manner. 
These high-redshift LyC leakers demonstrate statistically blue UV-continuum slopes ($\beta$), which is consistent with their high escape fraction of LyC photons. We find that these high-redshift LyC leakers span a wide range of specific star formation rate ($\log (\mathrm{sSFR}$/yr) from -8.6 to -6.7). 
Ten of these \lycl\ are located on the star formation main sequence, instead of all being in the starburst mode.
The results indicate that intense bursts of star formation are not necessarily required for the leakage of LyC photons for galaxies at $z>3$.
\end{abstract}

\section{Introduction} \label{sec:intro}
The epoch of reionization (EoR) is the last major phase change of the Universe, which happens at the redshift range of $6\lesssim z \lesssim 10$ constrained by the Gunn-Peterson trough observed in QSO spectra \citep[e.g.,][]{Fan2006} and the CMB observations \citep[e.g.,][]{Planck2016}. 
During the EoR, most of the neutral hydrogen in the intergalactic medium (IGM) is ionized by Lyman continuum photons (i.e. LyC photons, $\lambda < 912$ \AA) emitted by certain sources in the early universe, such as quasars and young stars in star-forming galaxies \citep[e.g.,][]{Finkelstein2012b, Robertson2013, Madau2015, Dayal2020}.
Recent studies have shown that quasars contribute only a small fraction of ionizing photons required by reionization \citep[e.g.,][]{Shen2020, Jiang2022, Matsuoka2023}, leaving star-forming galaxies to be the most likely sources of LyC photons contributing to cosmic reionization \citep[e.g.,][]{Finkelstein2019, Mason2019, Naidu2020}.

Theoretically, the ionizing photon budget during the EoR requires an average escape fraction of LyC photons from galaxies to be about $20$\% \citep[e.g.,][]{ouchi2009, Robertson2013, Robertson2015}. However, in observation, it is unlikely to detect the ionizing emission of galaxies in the EoR directly due to the increasing absorption of the IGM with redshift. Therefore, we have to rely on lower redshift galaxies with leaking LyC radiation to study the mechanisms that drive the escape of LyC photons from galaxies. 

Low redshift analogs at $z\lesssim0.3$, owing to the negligible IGM absorption and the high sensitivity of HST/Cosmic Origins Spectrograph (COS), have been widely studied in search of diagnostics closely related to LyC escape \citep[e.g.,][]{Borthakur2014, Izotov2018a, Izotov2018b, Flury2022}. However, the low-redshift \lycl\ have complicated selection functions, and their properties may evolve to be distinct from galaxies in the EoR \citep[e.g.,][]{naidu2022}.

Direct detection of LyC emission becomes feasible only when redshift decreases to $z\lesssim4.5$, making $3<z<4.5$ the closest window to the EoR to study the mechanisms that drive the escape of LyC photons from galaxies. 

Since the 2000s, many efforts have been made to search for \lycl\ at $z > 3$ \citep[e.g.,][]{Steidel2001, Shapley2006, Iwata2009, Steidel2018, Vanzella2018, Marques-Chaves2021, Marques-Chaves2022}.
Till now, dozens of galaxies at $z>3$ have been identified as \lycl, indicating that some galaxies can have high escape fractions ($\gtrsim20\%$) of LyC photons \citep[e.g.,][]{Vanzella2012, Shapley2016, Vanzella2016, Yuan2021}.
On the other hand, stacking analyses of galaxies without LyC detections result in a relatively small escape fraction \citep[$\lesssim5\%$, e.g.][]{Grazian2016, Wang2023}. 
It may suggest the diversity of properties that are connected to the escape of ionizing photons.
Therefore, it is necessary to find the properties of \lycl\ that are connected to the LyC leakage, and then examine whether these properties are consistent with those of star-forming galaxies in the EoR where the LyC emissions are not detectable.

Several studies find that the LyC leakers at $z>3$ have blue UV slope $\beta$ and are starbursts \citep[e.g.,][]{Debarros2016,Begley2022,Pahl2023}, consistent with the results based on the low redshift \lycl\ \citep[e.g.,][]{Chisholm2022,Flury2022}. In contrast, some studies find no dependence of LyC leakage on the physical properties of galaxies \citep[e.g.,][]{Saxena2022, Liu2023}. A possible cause of the discrepancy is that these works are based on different samples and methods.

In this study, we focus on the \lycl\ in the GOODS South field \citep[GOODS-S,][]{Giavalisco2004} to take advantage of the rich data in this field. We attempt to include all LyC leakers at spectroscopic redshift $z>3$ detected in previous studies. 
We include \emph{Ion1}, first reported by \citet[][\citetalias{Vanzella2012}]{Vanzella2012} as a LyC leaker at $z=3.797$ based on the detection in the U-band with VIMOS \citep{LeFevre2003}. The LyC emission has been further confirmed by \citet[][\citetalias{Ji2020}]{Ji2020} using the F410M band image observed with the Hubble Space Telescope (HST).
\emph{Ion2} ($z=3.212$) is another \lycl\, whose LyC emission has been detected in the VIMOS spectrum, image, and HST F336W band \citep[][\citetalias{Debarros2016}]{Vanzella2015, Vanzella2016, Debarros2016}. 
Using the data of the Hubble Deep UV Legacy Survey \citep[HDUV,][]{Oesch2018}, \citet[][\citetalias{Yuan2021}]{Yuan2021} found a LyC leaker at $z=3.797$ (CDFS-6664), whose LyC emission has been detected in the HST F336W band. 
\citet[][\citetalias{Rivera-thorsen2022}]{Rivera-thorsen2022} carried out a bottom-up search using the UV images from the Ultraviolet Hubble Ultra Deep Field \citep[UVUDF,][]{Teplitz2013, Rafelski2015} and HDUV \citep{Oesch2018}. They identified two new \lycl\ at $z>3$. \citet[][\citetalias{Saxena2022}]{Saxena2022} reported 11 \lycl\ based on narrow band \citep{Guaita2016} and VIMOS-U band images \citep{Nonino2009}.
\citet[][\citetalias{Gupta2024}]{Gupta2024} newly discovered a \lycl\ (z19863) using the HST F336W filter. 
\citet[][\citetalias{Kerutt2024}]{Kerutt2024} identified 12 LyC leakers based on HDUV \citep{Oesch2018}.

In the above studies, different methods and parameters are employed, which makes it hard to make a fair comparison between different works.
Here we collect the information on all these LyC leakers and fit their spectral energy distributions (SEDs) with a common set of models to obtain their physical properties systematically. 

In this work, we use the multiwavelength data provided by 3D-HST catalog \citep{Skelton2014} and include the newest released data from James Webb Space Telescope \citep[JWST,][]{Gardner2023} to extend the wavelength coverage to the rest-frame near-infrared (NIR).
In the SED fitting, we include emission line models to alleviate the effect of strong emission lines on broadband fluxes \citep[e.g.,][]{Debarros2014, Yuan2019}.
Also, different from many previous works, we treat the escape fraction $f_{\rm esc}$ as a free parameter in the SED fitting. A few recent studies have already taken $f_{\rm esc}$ as a free parameter in their analysis \citep[e.g.,][]{Marques-Chaves2021, Marques-Chaves2022, Yuan2021, Kerutt2024}.
The escape fraction of ionizing photons is related to the number of young stars and therefore can affect the shape of the SEDs and the estimation of the star-formation rates (SFRs) of galaxies \citep{Zackrisson2013, Rivera-thorsen2022}.

We introduce the sample of \lycl\ in Section \ref{sec:211samp}, and present the data and method used in our study in Section \ref{sec:method}. In Section \ref{sec:res}, we investigate the star-forming properties of the \lycl\ based on the results of SED fitting. Finally, we summarize the work in Section \ref{sec:sum}. We adopt a standard cosmology model with parameters $\Omega_{M} = 0.3$, $\Omega_{\Lambda} = 0.7$, $H_{0} = \rm{70\ km\ s^{-1} Mpc^{-1}}$. 
All magnitudes throughout this paper are given in the AB system.

\section{A Sample of LyC Leakers at $z>3$} \label{sec:211samp}
We have collected a sample of \lycl\ in the GOODS-S field with spectroscopic redshifts at $z>3$ from seven works \citepalias{Ji2020, Debarros2016, Yuan2021, Saxena2022, Rivera-thorsen2022, Gupta2024, Kerutt2024}.

We remove F336W-1041 in \citetalias{Rivera-thorsen2022} and CDFS 18454 in \citetalias{Saxena2022} because their SEDs cannot be fitted well ($\chi^2>5$) using their spectroscopic redshifts, which may be due to mismatched redshifts.
We exclude another source, CDFS 13385 in \citetalias{Saxena2022} because it is likely to be a low-redshift contaminator according to \citetalias[][]{Rivera-thorsen2022}. Furthermore, we find that CDFS 12448 in \citetalias{Saxena2022}, F336W-189 in \citetalias{Rivera-thorsen2022}, and 3052076 in \citetalias{Kerutt2024} are the same object.\cdfs in \citetalias{Yuan2021} and 122021111 in \citetalias{Kerutt2024} are also the same object.
The final sample contains 23 \lycl, as presented in Table~\ref{tab:lyc_sample}. These LyC leakers have an average redshift of about 3.505. Their redshifts range from 3.084 to 4.426.

We note that the LyC signals in this sample are detected in various bands with different signal-to-noise ratio (S/N) cuts.
As listed in Table \ref{tab:lyc_sample}, eight \lycl\ are detected in the LyC band with an S/N larger than 3. All the galaxies in our sample have detections more than 2$\sigma$ in the LyC band. Among these LyC leakers, \ionone, \iontwo, and F336W-189 are detected in two or more different bands.
In Table~\ref{tab:lyc_sample}, we summarize the basic information about the LyC signal of each source derived in previous works, including the bands in which the LyC signal was observed, the S/N of the LyC detection, and $f_\mathrm{esc}$ measurements. We also investigate the multiwavelength images of each object and record an offset if the separation between the centers of the LyC band and the nonionizing band is greater than 0.1\arcsec.

For comparison, we also include three high-confidence LyC leakers in other fields, which are \emph{Ion3} at $z \sim 4$, J0121+0025 at $z \sim 3.244$, and J1316-2614 at $z \sim 3.6130$ \citep{Vanzella2018, Marques-Chaves2021, Marques-Chaves2022, Marques-Chaves2024}.
No significant offset between the LyC and nonionizing UV emission is found in these LyC leakers.
The methodologies applied in these works are consistent with ours, allowing us to use their results on the physical properties directly.

\begin{center}
\begin{deluxetable*}{llllllrrr}[thb]
\tabletypesize{\scriptsize}
\tablewidth{0pt}
\tablecaption{Sample of LyC leakers and basic properties.
\label{tab:lyc_sample}}
\tablehead{\colhead{Name$^{a}$} &
\colhead{R.A.} &
\colhead{Decl.} &
\colhead{Redshift} &
\colhead{$f_\mathrm{esc}^{b}$} &
\colhead{LyC band} &
\colhead{$\rm S/N_{LyC}$} &
\colhead{Offset$^{c}$} & 
\colhead{Reference}}
\startdata
1181371$^\dag$ & 53.1358 & -27.7955 & 3.084 & $0.88 \pm 0.07$ & F336W & $5.01 \sigma$ & Yes & \citetalias{Kerutt2024} \\
$z19863^\dag$ & 53.1699 & -27.7684 & 3.088 & $0.24 \pm 0.06$ & F336W & $\sim 4 \sigma$ & Yes & \citetalias{Gupta2024} \\
$\iontwo^\dag$ & 53.013515 & -27.7552331 & 3.212 & $0.25^{+0.004}_{-0.000}$ & F336W (U, $\rm Spec_\mathrm{VIMOS}$) & $\sim 10 \sigma$ & No & \citetalias{Debarros2016} \\
109004028$^\dag$ & 53.0994 & -27.8392 & 3.267 & $0.34 \pm 0.10$ & F336W & $3.03 \sigma$ & Yes & \citetalias{Kerutt2024} \\
F336W-189$^\dag$ & 53.1679012 & -27.7979524 & 3.46 & $0.36 \pm 0.22$ & F336W (U) & $\sim 3.6 \sigma$ & Yes & \citetalias{Rivera-thorsen2022} \\
$\ionone^\dag$ & 53.0693219 & -27.7148184 & 3.794 & $0.05 \pm 0.02$ & U (F410M) & $\sim 10.3 \sigma$ & Yes & \citetalias{Ji2020} \\
$\rm \cdfs^\dag$ & 53.13885833 & -27.83537861 & 3.797 & $0.38 \pm 0.07$ & F336W & $\sim 5 \sigma$ & Yes & \citetalias{Yuan2021} \\
126049137$^\dag$ & 53.2042 & -27.8172 & 4.426 & $0.69 \pm 0.10$ & F336W & $4.39 \sigma$ & No & \citetalias{Kerutt2024} \\
\hline
CDFS 16444 & 53.1446546 & -27.7711112 & 3.128 & $0.30 \pm 0.07$ & NB3727 & $> 2 \sigma$ & No & \citetalias{Saxena2022} \\
1521589 & 53.1283 & -27.7887 & 3.152 & $0.79 \pm 0.15$ & F336W & $ 2.69 \sigma$ & Yes & \citetalias{Kerutt2024} \\
CDFS 24975 & 53.1457298 & -27.6869713 & 3.187 & $0.27 \pm 0.11$ & NB3727 & $> 2 \sigma$ & - & \citetalias{Saxena2022} \\
CDFS 9358 & 53.1247187 & -27.8245184 & 3.229 & $0.14 \pm 0.07$ & NB3727 & $> 2 \sigma$ & - & \citetalias{Saxena2022} \\
119004004 & 53.1891 & -27.8363 & 3.314 & $0.26 \pm 0.14$ & F336W & $2.59 \sigma$ & No & \citetalias{Kerutt2024} \\
CDFS 5161 & 53.0545613 & -27.862915 & 3.42 & $0.53 \pm 0.24$ & NB396 & $> 2 \sigma$ & Yes & \citetalias{Saxena2022} \\
CDFS 15718 & 53.1595397 & -27.7767456 & 3.439 & $0.28 \pm 0.10$ & NB396 & $> 2 \sigma$ & No & \citetalias{Saxena2022} \\
CDFS 19872 & 53.1995306 & -27.7414921 & 3.452 & $0.29 \pm 0.30$ & NB396 & $> 2 \sigma$ & - & \citetalias{Saxena2022} \\
CDFS 9692 & 53.0268389 & -27.8209671 & 3.47 & $0.38 \pm 0.19$ & NB396 & $> 2 \sigma$ & Yes & \citetalias{Saxena2022} \\
CDFS 20745 & 53.0459052 & -27.7336266 & 3.495 & $0.38 \pm 0.30$ & NB396 & $> 2 \sigma$ & No & \citetalias{Saxena2022} \\
3452147 & 53.1541 & -27.7988 & 3.521 & $0.47 \pm 0.14$ & F336W & $2.60 \sigma$ & Yes & \citetalias{Kerutt2024} \\
4062373 & 53.1792 & -27.7829 & 3.663 & $0.74 \pm 0.13$ & F336W & $2.21 \sigma$ & Yes & \citetalias{Kerutt2024} \\
4172404 & 53.1851 & -27.7839 & 3.672 & $0.31 \pm 0.15$ & F336W & $2.88 \sigma$ & Yes & \citetalias{Kerutt2024} \\
5622786 & 53.1604 & -27.8174 & 4.005 & $0.53 \pm 0.11$ & F336W & $2.05 \sigma$ & Yes & \citetalias{Kerutt2024} \\
122032127 & 53.1326 & -27.8374 & 4.348 & $0.77 \pm 0.14$ & F336W & $2.33 \sigma$ & Yes & \citetalias{Kerutt2024} \\
\enddata
\tablenotetext{^\dag}{The high confidence sources.}
\tablenotetext{$a$}{The ID of \lycl\ reported by S22 is from the CANDELS catalog \citep{Guo2013}. The ID of \cdfs\ is from the GOODS-MUSIC catalog \citep{Grazian2006, Santini2009}.}
\tablenotetext{$b$}{The escape fraction of \iontwo\ is determined using the relative escape fraction from \citetalias{Debarros2016} and the E(B-V) values derived from SED fitting (see Section~\ref{sec:method})}
\tablenotetext{$c$}{An offset is noted if the separation between the centers of LyC band and the nonionizing band is greater than 0\arcsec.1.}
\end{deluxetable*}
\end{center}

\section{Data and Method} \label{sec:method}
In the GOODS-S field, there are deep photometric observations of galaxies covering a wide range of wavelengths. 
All the sources in our sample have broadband flux measurements from HST, VLT/ISAAC, and \emph{Spitzer}. 
16 of 23 LyC leakers have been observed by JWST. 
The extensive photometric data allows us to perform more reliable SED fitting.
By fitting these broadband fluxes in a consistent way, we can then study the physical properties of these galaxies systematically. 
We summarize the photometric data and the models we use in SED fitting, as well as those used in previous studies in Table~\ref{tab:method}.

\begin{center}
\scriptsize
\begin{deluxetable*}{lcrrrcrrrcccc}
\tabletypesize{\scriptsize}
\tablewidth{0pt}
\setlength{\tabcolsep}{0.5pt}
\tablecaption{The photometric data and SED models
\label{tab:method}}
\tablehead{
\colhead{~Refs.~} &
\colhead{~Data~} &
\colhead {~$\rm N_{phot}$~} &
\colhead{~Wavelength range~} &
\colhead{~IMF~} & 
\colhead{~SPS Model~} &
\colhead{~SFH~} &
\colhead{~$Z\ (Z_{\odot})$~} &
\colhead{~~Extinction curve~~} & 
\colhead{~~Nebular Model~~} &
\colhead{~$f_{esc}$~} 
}
\startdata
\citetalias{Saxena2022} & HST, Ks, \emph{Spitzer} & 14 & 0.4-8 $\mu$m & KB02 & \citetalias{Bruzual2003} & ExpDec & 0-1 & C00 & F13 & No \\
\citetalias{Rivera-thorsen2022} & HST & 11 & 0.2-1.5 $\mu$m & KB02 & \citetalias{Bruzual2003} & Delayed & 0.2 & C00 & F17 & No \\
\citetalias{Yuan2021} & HST, \emph{Spitzer} & 12 & 0.3-24 $\mu$m & Salpeter & \citetalias{Bruzual2003} & Delayed & 0.02-1 & C00/CCM89 & I10 & Yes \\
\citetalias{Debarros2016} & HST, \emph{Spitzer} & 9 & 0.4-8 $\mu$m & Salpeter & \citetalias{Bruzual2003} & Constant & 0.2 & S79 & - & No \\
\citetalias{Ji2020} & HST, Ks, \emph{Spitzer} & 16 & 0.4-8 $\mu$m & K01 & FSPS & Delayed & 1 & C00 & F98,13 & No \\
\citetalias{Gupta2024} & HST, JWST & 18 & 0.4-4.8 $\mu$m & Chabrier & \citetalias{Bruzual2003} & Delayed & - & CF00 & - & No \\
\citetalias{Kerutt2024} & HST, Ks, \emph{Spitzer} & 16 & 0.3-24 $\mu$m & Chabrier & \citetalias{Bruzual2003} & Double exponential & 0.005-1 & C00/CCM89 & I10 & Yes\\
This work & ~~HST, Ks, \emph{Spitzer}, JWST~~ & 26 & 0.4-8 $\mu$m & Chabrier & \citetalias{Bruzual2003} & Delayed & 0.2 & C00/CCM89 & I10 & Yes
\enddata
\tablecomments{Column 1: Reference. 
Column 2: Photometric data sources. Column 3: Maximum number of photometric data points included in the SED fitting.
Column 4: The wavelength range of photometric data. Column 5: Assumed stellar initial mass function.
Column 6: Stellar population synthesis model used.
Column 7: Assumed star formation history.
Column 8: Assumed metallicity.
Column 9: Assumed extinction curve.
Column 10: Include nebular models in the SED fitting? (Yes/No)
Column 11: Set ionizing photon escape fraction as a free parameter. (Yes/No)
}
\tablenotetext{ }{Reference: [1] KB02: \citet{Kroupa2002}; [2] Salpeter: \citet{Salpeter1955}; [3] K01: \citet{Kroupa2001};
[4] FSPS: \citet{Conroy2009}; [5] C00: \citet{Calzetti2000}; [6]CCM89: \citet{Cardelli1989};[7] S89: \citet{Seaton1979}; [8] F13: \citet{Ferland2013}; [9] F17: \citet{Ferland2017};[10]I10: \citet{Inoue2010}; [11] F98: \citet{Ferland1998}; [12] F13: \citet{Ferland2013}; [13] Chabrier: \citet{Chabrier2003};[14] CF00: \citet{Charlot2000}}.
\end{deluxetable*}
\end{center}

\subsection{Photometric data} \label{subsec:photometric data}
The broadband fluxes of these sources are taken from the 3D-HST catalog \citep{Brammer2012, Skelton2014}, which includes the measurements from HST, VLT/IASSC and \textit{Spizter} (specifically, the HST F435W, F606W, F775W, F814W, F850LP, F125W, F140W, F160W bands; the VLT/IASSC $\rm K_{s}$ band; and the \textit{Spizter} IRAC1, IRAC2, IRAC3, IRAC4 bands). 
We use the total flux in the catalog introduced in \citet{Skelton2014}.

We take the JWST photometry for the 16 LyC leakers covered by the JWST Advanced Deep Extragalactic Survey \citep[JADES,][]{Eisenstein2023}, the First Reionization Epoch Spectroscopically Complete Observations \citep[FRESCO,][]{Oesch2023}, and the JWST Extragalactic Medium-band Survey \citep[JEMS,][]{Williams2023}.
We use the total fluxes in the JADES data release 2 catalog for their JWST photometry \citep[][]{Rieke2023}.

In the JADES catalog, CDFS 15718 is identified as two distinct objects, whereas it is listed as a single object in the 3D-HST catalog. 
In the SED fitting, we calculate the JWST flux for CDFS 15718 by summing the total fluxes of both counterparts.

\subsection{SED fitting} \label{subsec:sedfitting}
We use Code Investigating GALaxy Emission \citep[\texttt{CIGALE},][]{Boquien2019} to derive the physical properties of these \lycl\ by fitting their broadband flux to the total model incorporating stars, gas, and dust.
We note that \texttt{CIGALE} fitting is not suitable for the LyC bands because it takes an average IGM absorption \citep[][]{meiksin2006}.
We therefore exclude the LyC broadband flux (F225W, F275W, F336W) from the SED fitting.

We assume the Bruzual-Charlot Stellar Population Synthesis Models \citep[\citetalias{Bruzual2003},][]{Bruzual2003} and adopt the Chabrier initial mass function \citep[IMF, ][]{Chabrier2003}. We note that using a different IMF \citep[e.g., Salpeter][]{Salpeter1955} does not change our conclusion.
The star formation history is modeled using a delayed form, $SFR \propto t/\tau^2 \mathrm{exp} (t/\tau$), where $t$ is the age from the onset of star formation and $\tau$ is the characteristic time scale of star formation.
The metallicity of stars is set to 0.2$\mathrm{Z_{\odot}}$, which is a typical value for high redshift galaxies \citep[][]{Finkelstein2012}. 

For the stellar populations, we assume that the dust attenuation obeys the Calzetti Law \citep{Calzetti2000}, extended with the \citet{Leitherer2002} curve between the Lyman break and 1500 \AA, and the $E(B-V)$ ranges from $0.001$ to $0.15$. 
For the nebular emission, we assume a Milky Way extinction curve \citep{Cardelli1989}. We assume that the nebular $E(B-V)$ is equal to the stellar $E(B-V)$ \citep[e.g.,][]{yuan2018,buat2018}.

\texttt{CIGALE} models the nebular emission based on the templates of \citet{Inoue2011}, assuming that the nebular metallicity is equal to the stellar metallicity. The input parameters include the dimensionless ionization parameter $U$ ($U=q_\mathrm{ion}/c$, where $q_\mathrm{ion}$ is the ionization parameter) and the escape fraction of the LyC photons, $f_\mathrm{esc}$.
To compute the nebular emission spectrum, a normalized template of emission lines is selected based on the radiation field strength ($U$) and metallicity ($Z$). The spectrum is then rescaled to match the total ionizing photon luminosity for fitting. The ionizing photon luminosity is determined alongside the stellar population and corrected for the escape fraction of ionizing photons ($f_{\rm esc}$) using the following factor from \citet{Inoue2011}:
\begin{equation}
k=\frac{1-f_\mathrm{esc}-f_\mathrm{dust}}{1+\alpha_{1}(T_e)/\alpha_B(T_e)\times(f_\mathrm{esc}+f_\mathrm{dust})},
\end{equation}
where $\alpha_B$ is the case B recombination rate, $\alpha_1$ is the recombination rate to the ground level, $T_e$ the electron temperature in K. In the fitting, we take $T_e=10^4$ K. The corresponding $\alpha_B$ and $\alpha_1$ are set to be $2.58\times10^{-19} $m$^3$ s$^{-1}$ and $1.54\times10^{-19}$ m$^3$ s$^{-1}$, respectively (\citealt{Ferland1980}, see also \citealt{Boquien2019}). $f_\mathrm{dust}$ is the partial absorption by dust before ionization. Here we assume that the dust affects only the nonionizing UV so that $f_\mathrm{dust}$ in the equation is 0 (see e.g., Steidel et al. 2018). We assume $\log U = -2.5$ considering that this value corresponds to log$q_{ion}/\rm{cm\ s^{-1}}= 8.0$, which is consistent with the value for local LyC leakers \citep[][]{Nakajima2014}.

\begin{figure*}
 \centering
 \includegraphics[width=\textwidth]{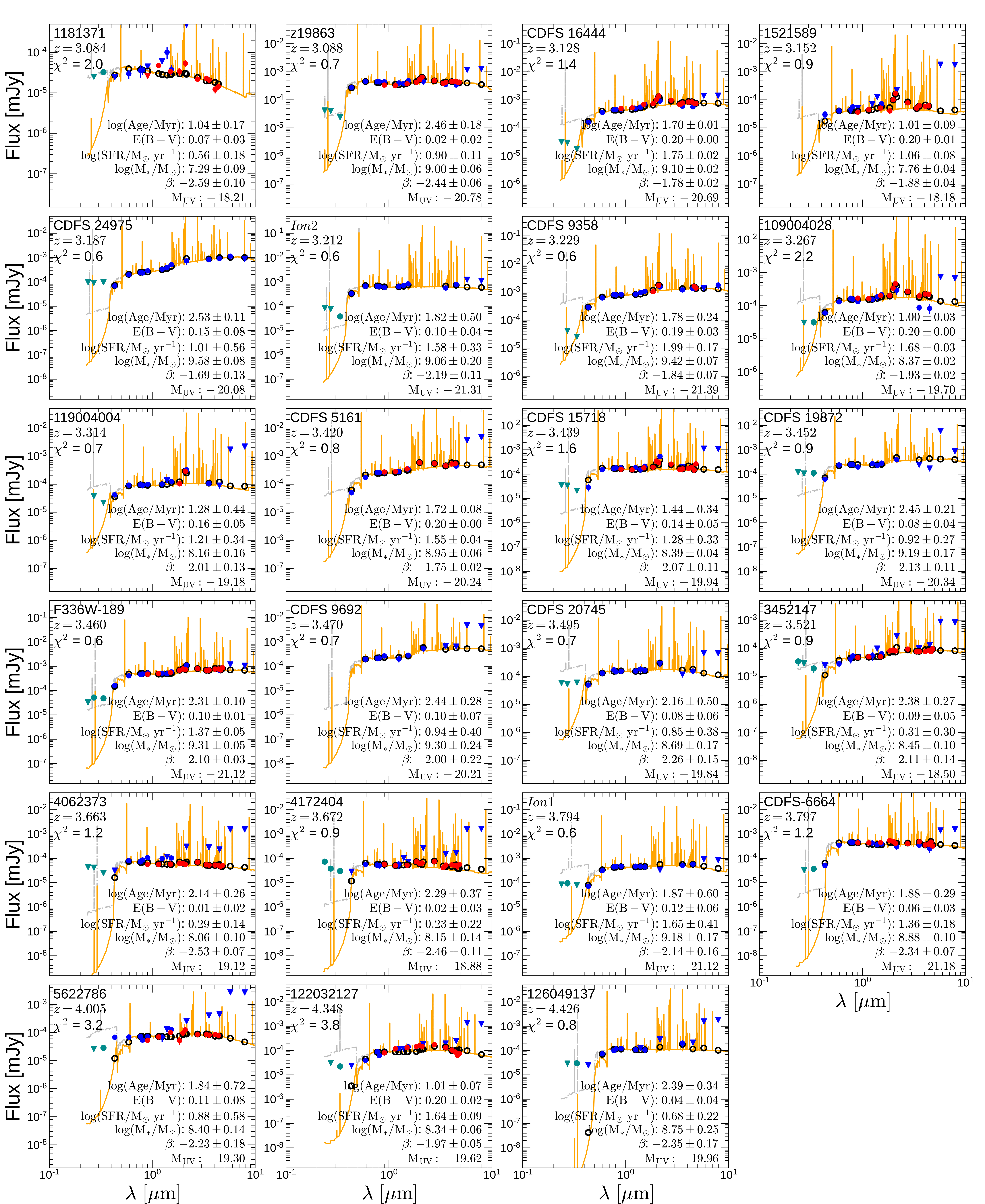}
 \caption{Observed SEDs (dots) of the LyC leakers. The red dots are measurements from the JADES catalog, and the blue and green dots are measurements from the 3D-HST catalog. The triangles indicate the $3\sigma$ upper limits for non-detections. We exclude photometric points (in green) affected by the IGM. We show the best SED fit with orange solid lines (IGM attenuated) and grey dashed lines (IGM unattenuated). The integrated fluxes based on the best SED fit are shown in black circles.
 }
 \label{fig:sed}
\end{figure*}

\begin{figure*}
 \centering
 \includegraphics[width=0.95\textwidth]{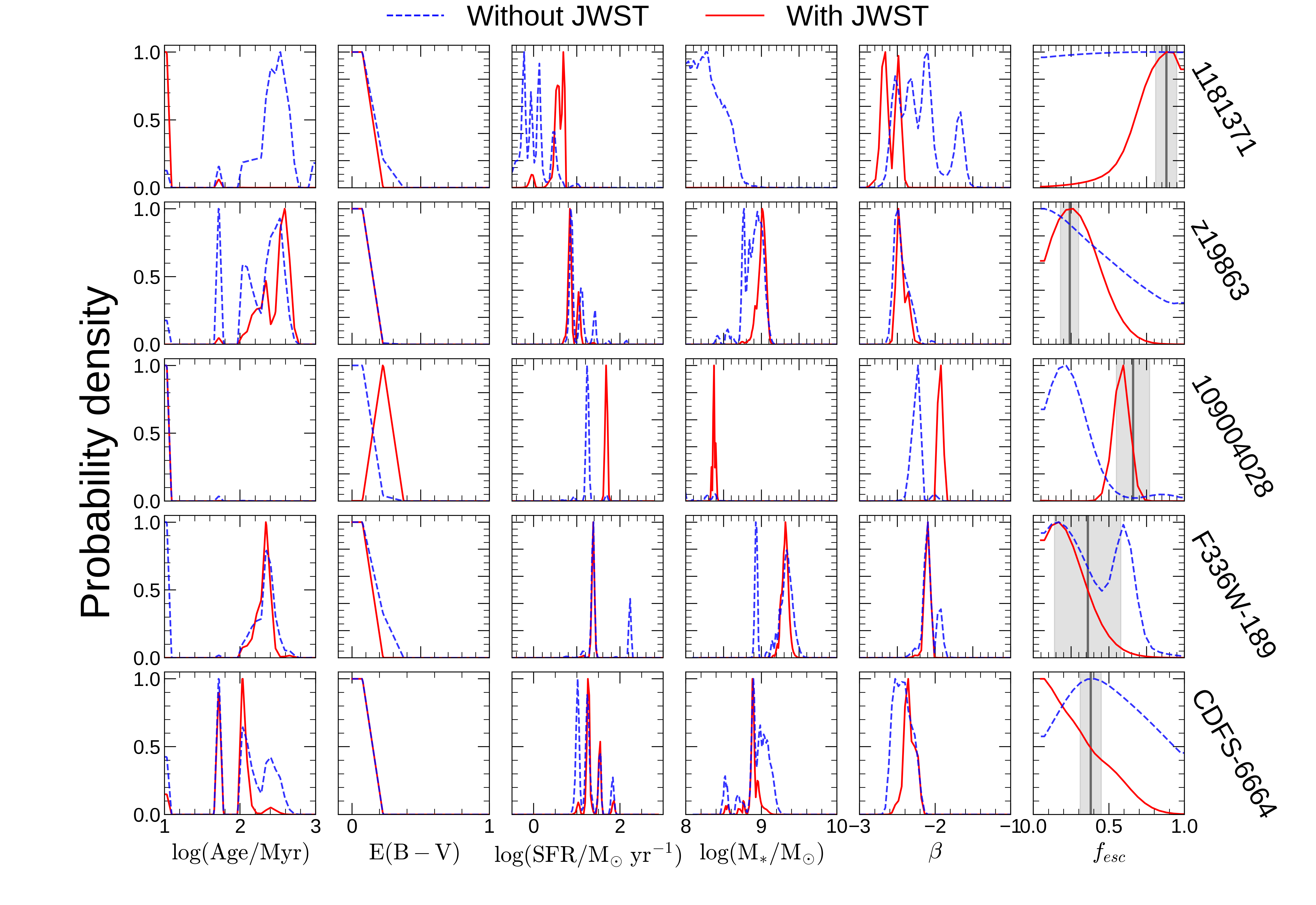}
 \caption{Probability density distributions for physical parameters (age, E(B-V), SFR, $M_*$, $\beta$) derived from SED fitting for 5 of the high confidence LyC leakers. The blue dashed lines are the probability distributions using 3D-HST photometric observations. The red solid lines are the probability distributions using 3D-HST photometric observations plus the observations of JWST. 
 The use of the JWST photometric data provides better constraints on these star-forming parameters. In the last column, we also present the probability density distributions for $f_\mathrm{esc}$. For reference, the $f_{\rm esc}$ values with their 1 $\sigma$ errors reported in the literature are represented by the black lines and the grey areas. 
 }
 \label{fig:prop}
\end{figure*}

We compare the photometric data and the models used in our SED fitting with previous works in Table \ref{tab:method}. Our SED fitting includes more photometric data to constrain the physical properties of the \lycl. Considering that the escape fraction of ionizing photons of these \lycl\ should be significantly larger than zero, it is more reasonable to set $f_{\rm esc}$ as a free parameter than to assume a zero escape fraction. Since we use a common set of photometric data and model assumptions for these \lycl\, the physical properties of these galaxies can be studied systematically. 

The output parameters are estimated using a Bayesian approach.
The parameters and the corresponding uncertainties are thus estimated from the probability distribution function (PDF) by taking the likelihood-weighted mean and the standard deviation of all models. 
The goodness of the fit is estimated using the reduced $\chi^2$ of the best model.
Figure~\ref{fig:sed} shows the results of the SED fitting. 19 out of 23 LyC leakers in our sample have $\chi^2<2.0$.

For 15 of the 16 sources observed by JWST, more than four bands have been included in the SED fitting (see Figure \ref{fig:sed}), offering improved constraints on the physical parameters compared to those without the JWST bands, particularly for the stellar masses, as shown in Figure~\ref{fig:prop}.

\section{Results} \label{sec:res}
In this section, we examine the star-forming properties of these \lycl.
We focus on the $\rm \beta$-$M_\mathrm{UV}$ and SFR-$M_{*}$ relations of the high-redshift \lycl\ and compare them with the results derived from normal star-forming galaxies at a similar redshift. 
We present our results in Figure~\ref{fig:beta-m&ms-sfr}, where the color bar encodes $f_{\rm esc}$ of LyC leakers.
We note that the $f_{\rm esc}$ calculated here strongly depends on the IGM models, and therefore has considerable uncertainties.

\subsection{$\beta$-$\rm M_{\rm UV}$ relation}
\label{subsec:uvprop}
The UV-continuum spectrum slope $\beta$ ($f_{\lambda}\propto \lambda^{\beta}$) is an important parameter as it is sensitive to the metallicity, age, and especially the dust content within a galaxy. Since nebular continuum emission is significantly redder than the continuum emission from young and low-metallicity stars \citep[e.g.,][]{Leithererheckman1995}, $\beta$ can also reflect the strength of nebular emission in a galaxy. Consequently, $\beta$ can set constraints on the escape fraction of LyC photons \citep[e.g.,][]{dunlop2013, Zackrisson2013}. 

\begin{figure*}[th]
\centering
\includegraphics[width=\textwidth]{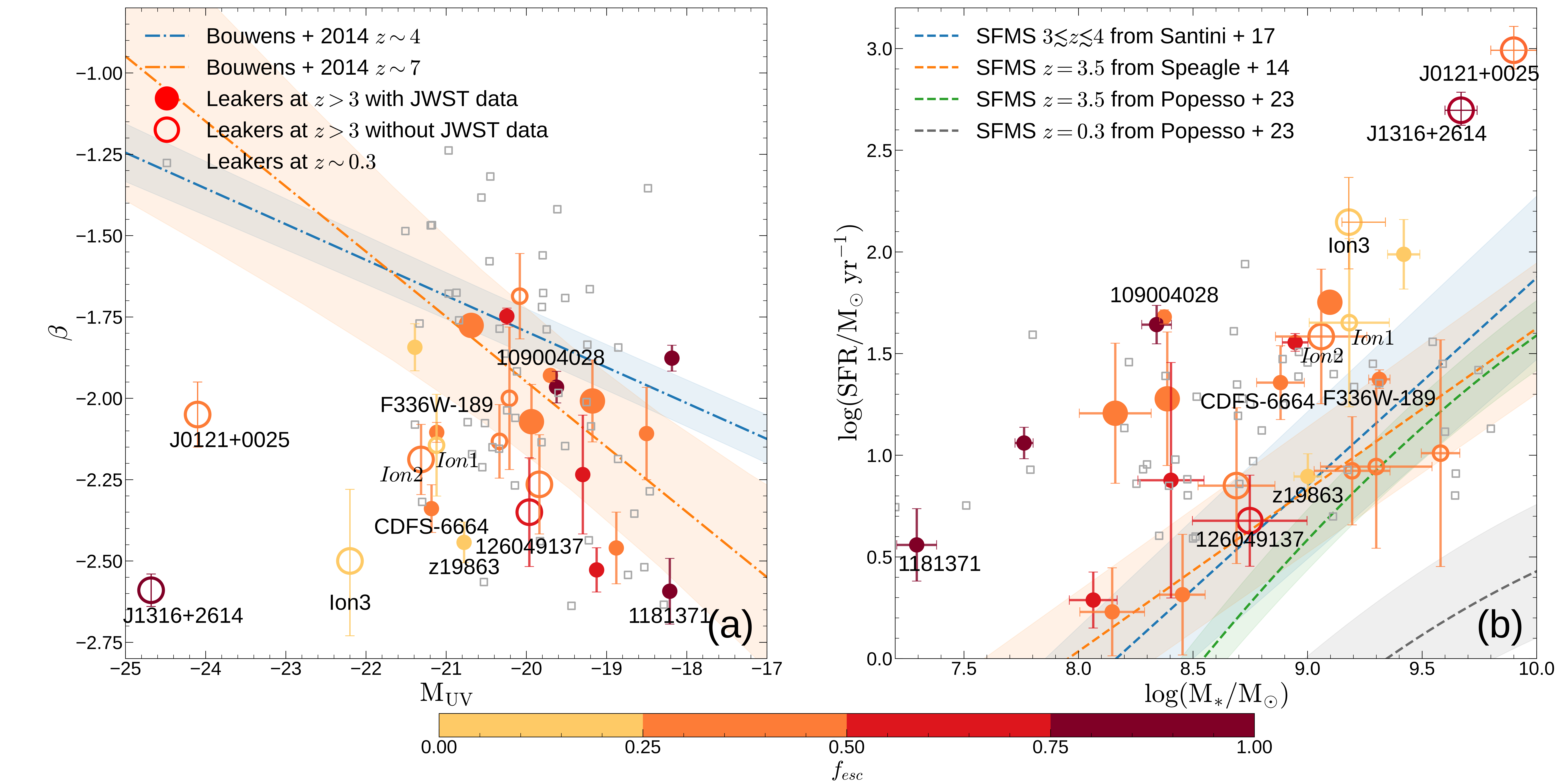}
\caption{
Star-forming properties of LyC leakers at $z>3$. The filled and open circles indicate \lycl\ with and without JWST photometric data, respectively.
 Three LyC leakers at $z>3$ in other fields (Ion3, J0121+0025, J1316+2614) are also presented in this figure.
The color indicates the $f_{esc}$ of each LyC leaker.
LyC leakers with offset (see Table \ref{tab:lyc_sample}) are indicated by small symbols.
For comparison, we over plot a low-z ($z\sim0.3$) sample of LyC leakers from \citet{Flury2022} as squares.
(a) UV-continuum slope $\beta$ as a function of UV absolute magnitude $M_{\rm UV}$. 
The blue and orange lines show the $\beta$-$M_{\rm UV}$ observed at $z \sim 4$ and $7$ from \citet{Bouwens2014}, respectively. The shaded areas represent the 1-$\sigma$ dispersion of these relations.
Most \lycl\ have bluer $\beta$ than the star-forming galaxies at similar redshifts ($z\sim4$). Nine \lycl\ have extremely blue $\beta$, which are located even below the relation at $z \sim 7$. 
The $z\sim7$ relation is also applicable for low-redshift ($z\sim0.3$) \lycl\, as reported by \citet{Chisholm2022}. 
(b) SFR as a function of stellar mass.
The blue, orange, and green dashed lines indicate the star formation main sequence (SFMS) at $z \sim 3.5$ derived from \citet{Santini2017}, \citet{Speagle2014}, and \citet{Popesso2023}, respectively.
The gray dashed line indicates the SFMS at $z\sim 0.3$ derived from \citet{Popesso2023}.
Shaded areas represent the 1-$\sigma$ dispersion of these relations.}
\label{fig:beta-m&ms-sfr}
\end{figure*}

We investigate the relation of $\beta$ and $\rm M_{\rm UV}$ of the $z>3$ \lycl\ in our sample (Figure \ref{fig:beta-m&ms-sfr}a). The UV-continuum slope $\beta$ is computed by fitting a straight line to the $F_{\lambda}$ spectrum in log-log space over the wavelength ranges defined in Table 2 of \citet{Calzetti1994}. $\rm M_{UV}$ is derived from the flux density at 1500 \AA\ of the best-fit SED. 

Figure \ref{fig:beta-m&ms-sfr}a shows the UV-continuum slope $\beta$ and $\rm M_{\rm UV}$ for the high-redshift \lycl. 
The $\rm M_{UV}$ of the high-redshift sample ranges from -18.2 mag to -21.39 mag. The UV slope ranges from -2.6 to -1.7.
We do not find a clear trend that the \lycl\ with lower UV luminosities have bluer UV slopes, as indicated in studies for high-redshift star-forming galaxies or low-redshift LyC leakers \citep[e.g.,][]{Bouwens2014, Chisholm2022}. This may be partly due to the small size of the sample.

Most of the LyC leakers lie significantly below the mean relation at $z\sim4$ and show very blue $\beta$ slopes, including seven galaxies in the GOODS-S (\ionone, \iontwo, CDFS-6664, F336W-189, z19863, 1181371, and 126049137) which are either detected in more than one observation or with high signal-to-noise ratios ($>3$) and thus are the most reliable LyC leakers in our sample.
The blue UV-continuum slopes of these galaxies suggest that these LyC leakers have young stellar populations, low metallicity, or low dust attenuation, consistent with the condition that allows LyC photons to leak. 

Based on a sample of low-redshift ($z\sim0.3$) LyC leakers, \citet{Chisholm2022} find that the $\beta$-$\rm M_{UV}$ relation for low-redshift LyC leakers is similar to that of $z\sim7$ galaxies predicted from \citet{Bouwens2014}. Figure~\ref{fig:beta-m&ms-sfr}a shows that LyC leakers at $z>3$ statistically have bluer $\beta$ slopes compared with $z\sim7$ LBGs or low-redshift LyC leakers.
Given that the high-redshift LyC leakers in our sample have an average of $f_\mathrm{esc=42\%}$, while the average $f_\mathrm{esc}$ of low-redshift LyC leakers is $8\%$ (range from $0.4\%$ to $58\%$), the observed bluer $\beta$ is consistent with the correlation between $\beta$ and $f_\mathrm{esc}$ as previously found by \citet{Chisholm2022}.

\subsection{SFR-$\rm M_{*}$ relation} \label{subsec:sfacti}
Figure~\ref{fig:beta-m&ms-sfr}b shows the SFR-$\rm M_{*}$ relation of the high-redshift \lycl\ in our sample. These high-redshift LyC leakers span a wide range of specific star formation rate ($\log (\mathrm{sSFR}$/yr) from -8.6 to -6.7).
Using the star formation main sequence (SFMS) derived from \citet{Speagle2014}, \citet{Santini2017}, and \citet{Popesso2023} as the reference at $z>3$, we find that 16 LyC leakers (including 3 LyC leakers not in GOODS-S) are distributed in the starburst region and the other 10 are on the SFMS.

The most vigorous starbursts in our sample, which have the
highest specific SFR (sSFR) and are farthest above the main
sequence lines in Figure 3b, show high escape fractions of LyC photons.
Among them, four LyC leakers exhibit extremely high escape fractions
($f_\mathrm{esc} > 75\%$). However, we do not see a clear trend of $f_\mathrm{esc}$ as a function of sSFR. In fact, the average $f_{\rm esc}$ of LyC leakers on the SFMS and those in the starburst region are comparable.

Theoretically, vigorous starbursts generate a large number of ionizing photons and create density-bound HII regions or optical thin channels for the ionizing photons to escape.
However, the fact that there are ten \lycl\ on the SFMS indicates that the extreme star formation may not be the fundamental reason that causes the escaping of ionizing photons. 

The data suggests that starburst activity is not a prerequisite for high-redshift galaxies ($z>3$) to have a high escape fraction of ionizing photons. This finding contrasts with the characteristics of low-redshift LyC leakers in \citet{Flury2022}. As shown in Figure \ref{fig:beta-m&ms-sfr}b, all the low-redshift LyC leakers are identified as starbursts relative to the star-forming main sequence at $z\sim0.3$. This distinction between high- and low-redshift LyC leakers may imply that different mechanisms are at play in the escape of LyC photons from these galaxies. Notably, LyC leakers at both $z>3$ and $z\sim0.3$ are positioned similarly on the SFR-$M_{*}$ plot, indicating a comparable level of star formation activity.

Some of our galaxies show offset between their LyC and nonionizing emission, which may indicate possible foreground contaminations \citep{nestor2013}. 
In Figure \ref{fig:beta-m&ms-sfr}, we use smaller symbols to distinguish sources with LyC offsets from those without. We find that LyC leakers with offset distribute similarly to those without offset. Therefore, including sources with offset LyC does not affect our conclusion.

We will explore more about the physical properties that are shared by \lycl. Especially, we find that all of these high-redshift \lycl\ either show merging signatures or present an offset between the LyC emission and the optical emission which may suggest unresolved merger systems (Yuan et al. \textit{in press}). The interaction may perturb the ISM in galaxies and produce channels with low optical depth for LyC photons to escape \citep{Rauch2011, Gupta2024}. We will study in more detail the morphology of LyC leakers in another work, to further investigate the driving mechanisms of the escape of the ionizing photons (Zhu et al., \textit{in prep.}).

\section{Summary} \label{sec:sum}
We collect a sample of \lycl\ at $z>3$ in the GOODS-S field discovered by previous studies to investigate their physical properties systematically.
We use a uniform set of photometric data and apply a common set of models and parameters in the SED fitting.
We include the newest released JWST data in the analysis to give better constraints on the stellar mass and SFR. The \texttt{CIGALE} code also allows us to include the nebular emission models and adjust the $f_{\rm esc}$ parameters to improve the quality of fitting for \lycl.

Analyzing the derived properties, we find that the UV-continuum slope $\beta$ of high-redshift LyC leakers is statistically bluer than that of star-forming galaxies at the same redshift. Furthermore, the slope is bluer than low-redshift LyC leakers. The result is consistent with previous studies indicating a correlation between a bluer $\beta$ and an increased $f_\mathrm{esc}$.

Our analysis also reveals that 10 out of the 23 high-redshift LyC leakers do not fall into the starburst category. These 10 galaxies align with the properties of main sequence galaxies at similar redshifts. This result is in contrast to the findings for low-redshift ($z \sim 0.3$) LyC leakers, which uniformly display significant starburst activity at $z\sim0.3$.

Our results indicate that the intense bursts of star formation are not necessarily required for the LyC leakage at high redshift ($z>3$). In another work, we will further investigate the properties that may be related to the escape of LyC photons (Zhu et al. \textit{in Prep.}).

Compared with previous works, we include more LyC leakers and estimate the properties in a consistent way. By integrating all the \lycl\ from previous works, we have obtained more information on the physical properties of LyC leakers. Our study is still limited by the small sample size of the high-redshift \lycl. The upcoming China Space Station Telescope \citep[CSST,][]{Zhan2018} is equipped with the Multi-Channel Imager (MCI), which will carry out extremely deep observations in UV and optical bands with a spatial resolution comparable to HST and therefore will be able to detect more high-redshift \lycl. 
Furthermore, JWST will release more infrared data. 
With the aid of these future observations, we may further examine our conclusions using a larger sample with better data quality.

\begin{acknowledgments}
We thank the referee for providing helpful comments that have improved this work.
This work is supported by the National Key R\&D Program of China No.2022YFF0503402. It is partly supported by the Funds for Key Programs of Shanghai Astronomical Observatory. FTY acknowledges support from the Natural Science Foundation of Shanghai (Project Number: 21ZR1474300). ZYZ acknowledges support by the National Science Foundation of China (12022303), and the China-Chile Joint Research Fund (CCJRF No. 1906). We also acknowledge the science research grants from the China Manned Space Project, especially, NO. CMS-CSST-2021-A04, CMS-CSST-2021-A07. 

This work is based on observations taken by the 3D-HST Treasury Program (GO 12177 and 12328) with the NASA/ESA HST, which is operated by the Association of Universities for Research in Astronomy, Inc., under NASA contract NAS5-26555.

This work is based on observations made with the NASA/ESA/CSA James Webb Space Telescope. The data were obtained from the Mikulski Archive for Space Telescopes at the Space Telescope Science Institute, which is operated by the Association of Universities for Research in Astronomy, Inc., under NASA contract NAS 5-03127 for JWST. 

All the data used in this paper can be found in MAST: \dataset[10.17909/T9JW9Z \citep{https://doi.org/10.17909/t9jw9z}]{http://dx.doi.org/10.17909/T9JW9Z}, \dataset[10.17909/8tdj-8n28 \citep{https://doi.org/10.17909/8tdj-8n28}]{http://dx.doi.org/10.17909/8tdj-8n28} and 
\dataset[10.17909/fsc4-dt61 \citep{https://doi.org/10.17909/fsc4-dt61}]{https://doi.org/10.17909/fsc4-dt61}.

\end{acknowledgments}

\vspace{5mm}
\facilities{HST (ACS, WFC3), VLT (IASSC), Spizter (IRAC), JWST (NIRCam)}

\software{Astropy \citep{2013A&A...558A..33A,2018AJ....156..123A,2022ApJ...935..167A}, CIGALE \citep{Boquien2019}, Numpy \citep{harris2020array}, Matplotlib \citep{Hunter:2007}}

\clearpage
\bibliography{main}{}
\bibliographystyle{aasjournal}
\end{CJK*}
\end{document}